\documentclass[proceedings]{JHEP}

\usepackage{amsmath}
\usepackage{amsthm}
\usepackage{amsfonts}
\usepackage{amssymb}

\theoremstyle{definition}

\theoremstyle{remark}

\newcommand{\beq}{\begin{equation}}
\newcommand{\eeq}{\end{equation}}
\newcommand{\USL}{\CU_q(\fsl(2,\BR))}

\newcommand{\pa}{\partial}
\newcommand{\ot}{\otimes}
\newcommand{\ra}{\rightarrow}

\newcommand{\ti}{\times}

\newcommand{\fr}[2]{{\textstyle \frac{#1}{#2} }}

\newcommand{\fsl}{{\mathfrak s}{\mathfrak l}}

\newcommand{\bra}{\langle}
\newcommand{\ket}{\rangle}
\def\ew{\hspace*{-.5mm}}   \def\ppe{\hspace*{-1mm}}

\newcommand{\SJS}[6]{ \displaystyle{\bigl\{ } \ew 
\begin{array}{ll} {\scriptstyle #1 }
  \ppe & {\scriptstyle #2} \ppe \\[-2mm] {\scriptstyle #3}\ppe &
  {\scriptstyle #4}\ew \end{array}\big| \ew
\begin{array}{l} {\scriptstyle #5 }
  \ppe \\[-2mm] {\scriptstyle #6}\ew  \end{array}\displaystyle{\bigr\}_b}} 
\newcommand{\SJSP}[6]{ \displaystyle{\bigl\{ } \ew 
\begin{array}{ll} {\scriptstyle #1 }
  \ppe & {\scriptstyle #2} \ppe \\[-2mm] {\scriptstyle #3}\ppe &
  {\scriptstyle #4}\ew \end{array}\big| \ew
\begin{array}{l} {\scriptstyle #5 }
  \ppe \\[-2mm] {\scriptstyle #6}\ew  \end{array}\displaystyle{\bigr\}_b'}} 
 
\newcommand{\Fus}[6]{F_{{\scriptstyle #1}{\scriptstyle #2}}
  \hspace*{.3mm}\displaystyle{[} \ew \begin{array}{ll} {\scriptstyle #3 }
  \ppe & {\scriptstyle #4} \ppe \\[-2mm] {\scriptstyle #5}\ppe &
  {\scriptstyle #6}\ew \end{array}\displaystyle{]}}

\newcommand{\BOC}[6]{\,\displaystyle{[}\ew 
  \begin{array}{lll} 
  {\scriptstyle #1} \ppe
  & {\scriptstyle #3} \ppe & {\scriptstyle #5} \ew \\[-2mm] {\scriptstyle
  #2} \ppe & {\scriptstyle #4}\ppe & {\scriptstyle #6} \ew
  \end{array}\displaystyle{]}}

\newcommand{\al}{\alpha}
\newcommand{\be}{\beta}
\newcommand{\ga}{\gamma}
\newcommand{\Ga}{\Gamma}
\newcommand{\de}{\delta}
\newcommand{\De}{\Delta}

\newcommand{\Om}{\Omega}
\newcommand{\si}{\sigma}

\newcommand{\bL}{\bar{L}}

\newcommand{\bz}{\bar{z}}

\newcommand{\CF}{{\mathcal F}}

\newcommand{\CH}{{\mathcal H}}

\newcommand{\CL}{{\mathcal L}}

\newcommand{\CO}{{\mathcal O}}

\newcommand{\CR}{{\mathcal R}}
\newcommand{\CS}{{\mathcal S}}

\newcommand{\CU}{{\mathcal U}}
\newcommand{\CV}{{\mathcal V}}

\newcommand{\CZ}{{\mathcal Z}}

\newcommand{\BR}{{\mathbb R}}
\newcommand{\BD}{{\mathbb D}}

\newcommand{\BC}{{\mathbb C}}
\newcommand{\BF}{{\mathbb F}}

\newcommand{\BS}{{\mathbb S}}

\newcommand{\BZ}{{\mathbb Z}}

\DeclareMathOperator{\Tr}{Tr}

\newcommand{\rf}[1]{(\ref{#1})}

\newcommand{\aufz}
{\begin{list}{$\bullet$}{\topsep0cm \itemsep0cm \parsep0cm}}
\newcommand{\eaufz}{\end{list}}
\newcounter{num}
\newcommand{\remlst}{\begin{list}
{(\arabic{num})}{\usecounter{num}\topsep0cm \itemsep0cm \parsep0cm}}
\conference{Nonperturbative Quantum Effects 2000}
\title{Remarks on 
Liouville theory with boundary}
\author{J.Teschner\\Institut f\"ur theoretische Physik,
Freie Universit\"at Berlin,\\Arnimallee 14, 14195 Berlin, Germany
\\e-mail: teschner@physik.fu-berlin.de}
\abstract{The bootstrap for Liouville theory with
conformally invariant boundary conditions will be discussed. 
After reviewing some results on one- and boundary two-point
functions we discuss some analogue of the Cardy condition linking
these data. This allows to determine the spectrum of the theory
on the strip, and illustrates in what respects the bootstrap for
noncompact conformal field theories with boundary is richer 
than in RCFT. We briefly indicate some connections with
$\USL$ that should help completing the bootstrap.}
\begin{document}
\section{Motivation}
D-branes on compact spaces can be studied in the ``stringy regime''
($\al'\sim\CO(1)$) by means of conformal field theory
in the presence of boundaries \cite{RS}\cite{FS}. The treatment
of D-branes on noncompact spaces requires consideration of 
CFT with continuous spectrum of Virasoro representations 
(noncompact CFT). Liouville theory may be considered as a
prototypical example of such CFT. It seems to be the
natural starting point for the development of techniques for
the exact study of the class of CFT that describe D-branes on noncompact
backgrounds. 
Moreover, physically interesting
examples such as the $SL(2)/U(1)$ black
hole or $AdS_3$ are closely related to Liouville theory
from the technical point of view.
\section{Liouville theory w/o boundary}

We will very briefly assemble a few facts concerning 
Liouville theory with periodic boundary conditions that
will be referred to later.

The classical 2D field theory is defined on $\BR\ti S^1$
by the Lagrangian
\[ 
\CL \;\,= \;\, \frac{1}{4\pi}(\pa_a\phi)^2\;+ \;\mu e^{2b\phi}.
\]
The spectrum is believed \cite{BCGT} to be of the following form:
\[ 
\CH\;=\; \int\limits_{\BS}d\al\; \CV_{\al}\ot\CV_{\al},\qquad
\BS=\frac{Q}{2}+i\BR^{+}
\]
where $\CV_{\al}$: hwr of Virasoro algebra (generators
$L_n$, $\bL_n$), highest weight $\De_{\al}=\al(Q-\al)$, $Q=b+\frac{1}{b}$.

Conformal symmetry 
allows one to map the cylinder to the complex plane 
via $z=e^{w}$.
Basic objects for the understanding 
of the theory are the local primary fields $V_{\al}(z,\bz)$,
\[ \begin{aligned}
{[}L_n,V_{\al}(z,\bz){]}=&z^n(z\pa_{z}+\De_{\al}(n+1))V_{\al}(z,\bz)\\
{[}L_n,V_{\al}(z,\bz){]}=&\bz^n(\bz\pa_{\bz}+\De_{\al}(n+1))V_{\al}(z,\bz),
\end{aligned} \]
Thanks to conformal symmetry they turn out to be 
fully characterized by their three point functions
\begin{equation} \label{threept}
D(\al_3,\al_2,\al_1)=
\bra 0|V_{\al_3}(\infty)V_{\al_2}(1)V_{\al_1}(0)|0\ket.
\end{equation}

A formula for $D(\al_3,\al_2,\al_1)$ has been
proposed by
Dorn, Otto and AL.B., A.B. Zamolodchikov \cite{DO,ZZ}.
The data $(\BS,D)$ can be seen to contain the full information 
about Liouville theory on $\BR\ti S^1$. In particular, it is possible
to reconstruct the Liouville field itself, to show that it satisfies
a proper generalization of the classical equation of motion, and to
verify the canonical commutation relations \cite{TR}.  

Let us note that 
the proposals concerning $(\BS,D)$  
are not rigorously proven so far. However, there exists ample
evidence to their support, cf. e.g. \cite{DO,ZZ,PT1,TR} 
and references therein.

\section{Liouville theory with boundary}
One needs to study the
quantization of the classical 2D field theory defined by the action
\[ \begin{aligned}
\CS \;\,= \;\, \int
\limits_{\rm S}d^2w\; & \Bigl( 
\frac{1}{4\pi}(\pa_a\phi)^2\;+ \;\mu e^{2b\phi}\Bigr)\\
& + \int
\limits_{\BR}d\tau \; 
\Bigl( \rho_1e^{b\phi}\big|_{\si=\pi}\; +\; \rho_2 e^{b\phi}
\big|_{\si=0}\Bigr),
\end{aligned} \]
where S: strip $\BR\ti[0,\pi]$.

Conformal symmetry is expected to organize $\CH$ as 
\begin{equation}\label{spec} 
\CH\;=\; \int\limits_{\BS^{\rm B}}d\al\; \,\CV_{\al},
\end{equation}
where $\CV_{\al}$: highest weight representation 
of the Virasoro algebra (generators
$L_n$) with highest weight 
$\De_{\al}=\al(Q-\al)$. The set $\BS^{\rm B}$ to be integrated over 
will in general depend on the values $\rho_2$, $\rho_1$ labelling the 
boundary conditions.

In discussions of the euclidean theory it is often convenient 
to map the strip to the upper half plane by means of the
conformal transformation $w\ra z=e^w$. 

One is mainly interested in 
primary fields $\CO_{\al}(z)$. 
One needs to consider two types of such fields:\\
$\bullet$ {\bf bulk} fields $V_{\al}(z)$, defined on 
$\{z\in\BC;\Im(z)>0\}$, which transform as
\[ \begin{aligned}
{[}L_n,\CO_{\al}(z){]}=& 
z^n(z\pa_z+\De_{\al}(n+1))\CO_{\al}(z)\\
+&\bz^n(\bz\pa_{\bz}+\De_{\al}(n+1))\CO_{\al}(z),
\end{aligned}\]$\bullet$ 
{\bf boundary} fields $\Psi^{\be}_{\rho_2\rho_1}(x)$,
$x\in\BR$:
\[ 
[L_n,\Psi^{\be}_{\rho_2\rho_1}(x)]=x^n(x\pa_x+\De_{\be}(n+1))
\Psi^{\be}_{\rho_2\rho_1}(x).
\]

\section{Bootstrap}
\subsection{Data}
The conformal Ward identitites together with
factorization of correlation functions 
lead to an unambigous construction of any 
correlation function in terms of a set of elementary
amplitudes (structure functions): This set contains in addition to the
three point function $D$ of the theory {\it without} boundary \rf{threept} 
the following data: 
\begin{itemize}
\item[$\bullet$] {\bf One point function}: 
\[
\left\langle V_{\alpha}(x)\right\rangle \;= A(\alpha|\rho) \;\left|
z-\bar{z}\right|  ^{-2\Delta_{\alpha}} 
\]
\item[$\bullet$] {\bf Boundary two point function}:
\[ \begin{aligned}
\bra  & \Psi^{\beta}_{\rho_{1}\rho_{2}}(x)
\Psi^{\beta'}_{\rho_{2}\rho_{1}}(y) \ket \;=\;\left|  x-y\right|
^{-2\Delta_{\beta}}\\[1ex]
& \quad\ti N_0\bigl(\de_{Q-\be,\be'}+ \de_{\be,\be'}
B(\beta|\rho_{1},\rho_{2})\bigr) 
\end{aligned} \]
\item[$\bullet$] {\bf Bulk-boundary two point function}:
\[ \begin{aligned}{} 
\bra  V_{\alpha}(z) & \Psi^{\beta}_{\rho\rho}(x)\ket \;=\\
& =\frac{A (\alpha,\beta
|\rho)}{\left|  z-\bar{z}\right|  ^{2\Delta_{\alpha}-\Delta_{\beta}}\left|
z-x\right|  ^{2\Delta_{\beta}}} 
\end{aligned}\]
\item[$\bullet$]
{\bf Boundary three point function}:
\[ \begin{aligned}{}  
\bra & \Psi^{\be_3}_{\rho_1\rho_3}(x_3) 
\Psi^{\be_2}_{\rho_3\rho_2}(x_2)\Psi^{\be_1}_{\rho_2\rho_1}(x_1)\ket
=\\[.2cm]
& =
\frac{C\BOC{\be_3}{\rho_3}{\be_2}{\rho_2}{\be_1}{\rho_1}}{\left|
x_{12}\right|^{-\Delta_{12}}\left|  x_{23}\right|^{-\Delta_{23}}
\left|  x_{31}\right|^{-\Delta_{31}}}, 
\end{aligned}\]
where $x_{ij}=x_i-x_j$, $\De_{ij}=\De_k-\De_i-\De_j$.
\end{itemize}
Two comments concerning the form of the boundary two point function 
seem to be in order: First note that the 
spectrum $\BS^{\rm B}$ will later be found 
to have both continuous and discrete parts in general, $\BS^{\rm B}=\BS\cup
\BD$. The symbol $\de_{\be,\be'}$ will accordingly be interpreted as 
a Kronecker-delta in the case that $\be,\be'\in\BD$, and as delta-distribution
in the case that $\be,\be'\in\BS$. We have furthermore left undetermined
a normalization factor $N_0$ since we will later find relations between
the normalizations of bulk- and boundary operators.
\subsection{Consistency conditions}
These data are restricted by consistency conditions similar
to those found by Cardy and Lewellen in the case of RCFT
\cite{C,CL,L}: Let us first note the two conditions
\begin{equation}\label{b-ass}
\begin{aligned}
\int\limits_{\BF_{21}^{\rm B}}d\be_s 
\Fus{\be_s}{\be_t}{\be_{3}}{\be_2}{\be_4}{\be_1}  & 
C\BOC{\be_4}{\rho_4}{\be_3}{\rho_3}{\be_s}{\rho_1}
\,C\BOC{\be_s}{\rho_3}{\be_2}{\rho_2}{\be_1}{\rho_1}\\
= \; & C\BOC{\be_4}{\rho_4}{\be_t}{\rho_2}{\be_{1}}{\rho_1}\,
 C\BOC{\be_t}{\rho_4}{\be_3}{\rho_3}{\be_2}{\rho_2},
\end{aligned} 
\end{equation}
expressing associativity of the OPE of boundary operators, and
\begin{equation}\label{VVb}
\begin{aligned}
\int\limits_{\BF_{21}}\frac{d\al}{B(\al)}\;   & 
D(\al,\al_2,\al_1)   A(\al|\rho)
\Fus{\al}{\be}{\al_{1}}{\al_1}{\al_2}{\al_2}
=\\
=& \frac{A(\al_2,\be|\rho)A(\al_1,\be|\rho)}{B(\be|\rho,\rho)}.
\end{aligned}
\end{equation}
The data appearing in \rf{b-ass}\rf{VVb}
that have not yet been introduced are the following: $B(\al)$ and 
$D(\al_3,\al_2,\al_1)$
are two-and three point functions in Liouville theory with   
periodic boundary conditions \cite{DO}\cite{ZZ}\footnote{The two point
function is recovered from the expression for $D$ by
sending $\al_3\ra 0$, cf. e.g. \cite{TR}}, $F_{\be\ga}[..]$ are 
the fusion coefficients that describe the relation between s-channel and
t-channel conformal blocks \cite{PT1}, the set $\BF_{21}$ is the set of 
$\al$ labelling the primary
fields $V_{\al}$ that appear in the OPE of two bulk fields $V_{\al_2}$ and
$V_{\al_1}$ \cite{TR} and $\BF_{21}^{\rm B}$ is similarly the set of 
$\be$ labelling the boundary fields $\Psi^{\be}_{\rho_3\rho_1}$
that appear in the OPE of $\Psi^{\be_2}_{\rho_3\rho_2}$ and 
$\Psi^{\be_1}_{\rho_2\rho_1}$. We consider $(D,F;\BF_{21})$ 
to be known from \cite{DO,ZZ,PT1,TR} and intend to determine
$(A,B,C;\BF_{21}^{\rm B})$ as solution of the consistency conditions 
linking these two sets of data.

In addition to the conditions written above there are only
two further conditions to consider \cite{L}: One comes 
from correlation functions of the form $\bra V\Psi\Psi\ket$ and will not 
be considered here. The final condition is an analogue of what
is known as the Cardy condition. It will be the main focus of this note.
However, in order to even formulate it, we will have to go through some
preparations. 
\section{Known structure functions}
\subsection{One point function}
The one point function $A(\al|\rho)$
has been determined in \cite{FZZ}\cite{TB}:
\[
\begin{aligned}
A\bigl(\fr{Q}{2}+iP|\rho\bigr)=
& A_0
\bigl(\pi \mu \ga(b^2)\bigr)^{-\frac{iP}{b}}\frac{\cos(4\pi sP)}{iP}
\cdot\\
& \cdot \Ga(1+2ibP)\Ga(1+2ib^{-1}P),
\end{aligned}\]
where $s$ parametrizes the boundary conditions via  
\[
\cosh\bigl(2\pi b s\bigr)=\frac{\rho}{\sqrt{\mu}}
\textstyle\sqrt{\sin(\pi b^2)}. 
\]
A couple of remarks are in order:\\[1ex]
{\bf 1.)} As opposed to \cite{FZZ}, 
we have included a prefactor $A_0$ independent of $\al$, $\rho$.
The Cardy condition that we intend to discuss will imply a relation 
between $A_0$ and the normalization factor $N_0$ that appears in the 
two point function. Moreover, condition \rf{VVb} further constrains the 
choice of $A_0$. We therefore expect all these quantities to be fixed
once the normalization of the bulk operators $V_{\al}$ is fixed. The
formula for $D$ that was proposed in \cite{ZZ} implies the 
following normalization:
\[
\bra V_{\frac{Q}{2}-iP'}(\infty)V_{\frac{Q}{2}+iP}(0)\ket_{\rm bulk}^{}
=2\pi\de(P'-P)
\]
for $P,P'\in\BR^+$.\\[1ex]
{\bf 2.)} $\rho$ determines $s$ only up to $s\ra s+inb^{-1}$. But 
$A(\al|s)$ is {\bf not} periodic under $s\ra s+inb^{-1}$.
We will therefore consider $s$ as true parameter for the
boundary conditions.
This is related to the existence of nontrivial quantum corrections 
to the effective action such as the appearance of a ``dual'' boundary
interaction $\tilde{\rho}e^{\phi/b}$.

Let us furthermore note that for real values of $\rho$ one finds 
two regimes: 
\begin{itemize}
\item[a)] $\rho\sqrt{\sin(\pi b^2)}>\sqrt{\mu}\quad\Rightarrow
\quad s\in\BR$,
\item[b)] $\rho\sqrt{\sin(\pi b^2)}<\sqrt{\mu}\quad\Rightarrow\quad
s\in i\BR$
\end{itemize}
{\bf 3.)} One may consider the one-point function as defining a boundary
``state'' ${}_B\bra s|$ as is done in RCFT with boundary \cite{C}:
\[ {}_{\rm B}^{}\bra s|\;=\; \frac{1}{2\pi}\int_{\BS}d\al \;A(\al|s)\;
{}_{\rm I}^{}\bra \al|
\]
where ${}_{\rm I}\bra \al|$: 
Ishibashi state constructed from the bulk-primary state 
$\bra P|$. In the present case one may note, however, that
the non-normalizability of ${}_{\rm B}^{}\bra s|$ is 
worse than usual due to the pole that $A(\al|s)$
has at $2\al=Q$. It will still be a useful object when considered as a 
distribution, analogous to the interpretation of the so-called
microscopic states in \cite{TR}.

\subsection{Boundary two point function}
The following expression was found in \cite{FZZ}:
\[\begin{aligned}
{} & B\bigl(\fr{Q}{2}+iP|s_2,s_1\bigr)\;=\\ 
&   
\bigl(\pi \mu \ga(b^2) b^{2-2b^2}\bigr)^{-\frac{iP}{b}}\;
\frac{\Ga_b(+2iP)}{\Ga_b(-2iP)}\cdot \\
& 
\frac{
 S_b\bigl(\fr{Q}{2}-i(P+s_1+s_2)\bigr)
 S_b\bigl(\fr{Q}{2}-i(P-s_1-s_2)\bigr)}
{S_b\bigl(\fr{Q}{2}+i(P+s_1-s_2)\bigr)
 S_b\bigl(\fr{Q}{2}+i(P+s_2-s_1)\bigr)}
\end{aligned}\]
Integral representations defining the special functions $G_b$ and 
$S_b$ can be found in \cite{FZZ}. Both of them are closely related to
the Barnes Double Gamma function \cite{Ba,Sh}, and $S_b$ was 
independently introduced under the name of ``quantum dilogarithm''
by L. Faddeev, cf. e.g. \cite{FKV} for related applications, properties
and references. 

Let us note that
\[ \boxed{\quad \bigl|B\bigl(\fr{Q}{2}+iP|s_2,s_1\bigr)\bigr|^2\;=\;1\quad}
\]
in the following three cases:\begin{enumerate}
\item $s_1,s_2\in\BR$,
\item $s_1,s_2\in i\BR$ and
\item $(s_1)^*=-s_2$.
\end{enumerate}

\subsection{Remark on uniqueness}

The above two results where obtained by considering special 
cases of the consistency conditions \rf{b-ass} and \rf{VVb} 
in which these equations reduce to finite difference equations
with coefficients that can be calculated by other means.
A proper discussion of uniqueness of solutions to these difference
equations is beyond the scope of the present note, but we would like
to mention that at present a strict proof of uniqueness 
requires assumptions concerning the 
analyticity of the structure functions 
w.r.t. $\al$, $\be$: Although invoking the $b\ra b^{-1}$-duality as 
in \cite{T1} yields uniqueness on any line parallel to the real axis
in the $\al$-plane, one needs further input to fix the dependence
on the imaginary part of $\al$, such as e.g. analyticity in certain
regions of the $\al$-plane. However, we do not believe that this
casts serious doubts on the correctness of the results obtained 
by these methods: On the one hand it is quite clear that 
alternative solutions would be of a rather bizarre kind,
and on the other
hand one may observe that analyticity w.r.t. $\al$ in certain
reqions of the complex plane is important for the physical 
consistency of Liouville theory, as will be discussed in the 
example of the bulk three point function $D$ in \cite{TR}.

\section{Remarks on canonical quantization on the strip}

In canonical quantization one would like to find a Hilbert space 
$\CH$ such that the algebra of fields is generated by
operators $\phi(\si,\tau)$ and $\Pi(\si,\tau)=(2\pi)^{-1}\pa_{\tau}
\phi(\si,\tau)$ that satisfy
\[ 
{[}\Pi(\si,\tau),\phi(\si',\tau){]}=-i\de(\si-\si').
\]
Time evolution should be generated by a Hamiltonian of the 
form
\begin{equation*}\begin{aligned}
H=\int\limits_0^{\pi}d\si & \;:\!\Bigl( 
\frac{1}{4\pi}\bigl((\pa_t\Phi)^2+(\pa_{\si}\Phi)^2\bigr)
\;+ \;\mu e^{2b\phi}\Bigr)\!:\\
+ &:\!\Bigl( \rho_le^{b\phi}\big|_{\si=\pi}\; +\; \rho_r e^{b\phi}
\big|_{\si=0}\Bigr)\!:.
\end{aligned}\end{equation*}
The dots are supposed to symbolize all normal orderings and
quantum corrections necessary to make $H$ well-defined 
\footnote{This may include addition of ``dual'' interactions 
like the boundary interaction $\tilde{\rho}e^{\phi/b}$.}. 

Assume that the Liouville zero mode $q\equiv \int_{0}^{\pi}d\si\phi(\si)$ 
can be diagonalized:
\[ \CH \;=\; \int\limits_{\BR}^{\oplus}dq \;\,\CH_{q}. \]
The zero mode $\Pi_0\equiv \int_0^{\pi}d\si \Pi(\si)$ then acts as
$\frac{1}{\pi i}\pa_{q}$ and 
\[
H=-\pa_q^2+\dots
\] 

The {\it exponential decay} of the interaction terms for $q\ra-\infty$
leads one to expect that 
\[ \CH_{q} \;\underset{q\ra-\infty}{\sim}\; \CF, \]
where $\CF$: Fock-space generated by the non-zero modes
$a_n$ acting on Fock-vaccuum $\Om$. One may therefore
characterize generalized eigenfunctions of $H$ 
by their asymptotic behavior for $q\ra-\infty$: 
\[ 
\Psi_E(q) \;\underset{q\ra-\infty}{\sim}\; 
 e^{iPq} F_N^+ + e^{-iPq}F_N^-,
\]
where $F_N^{\pm}\in\CF$, $N$: level, $E=\frac{Q^2}{4}+P^2+N$.

{\it Exponential blow-up} of interaction terms for $q\ra+\infty$ 
one the other hand leads one to expect {\it reflection}, i.e.
\[ F_N^-\; = \;\CR(P|s_2,s_1) F_n^+.\] 
Conformal symmetry determines $\CR(P|s_2,s_1)$ 
up to a scalar multiple $R(P|s_2,s_1)$. $R$ describes the asymptotic
behavior the wave-functions for primary states $|P,s_2,s_1\ket$ 
which satisfy
$L_n|P,s_2,s_1\ket=0$ for $n>0$, $L_0|P,s_2,s_1\ket=(P^2+\frac{Q^2}{4})
|P,s_2,s_1\ket$:
\[ 
\Psi_P(q) \;\;\underset{q\ra-\infty}{\sim}\;\;
\bigl( e^{iPq} + R(P|s_2,s_1)e^{-iPq}\bigr)\Om.
\]
One may observe that the {\rm reflection amplitude} $R(P|s_2,s_1)$
also describes the relation between $\Psi_P(q)$ and its
analytic continuation $\Psi_{-P}(q)$:
\[ 
\Psi_P(q)\; = \; R(P|s_2,s_1)\Psi_{-P}(q).
\] 

$R(P|s_2,s_1)$ is clearly a fundamental quantity for
describing the dynamics of Liouville theory on the strip.
It is therefore satisfactory to observe that  
{\it state-operator correspondence} leads to the
following
relation between $R(P|s_2,s_1)$ and the 
two point function $B(\be|s_2,s_1)$:
\[ R(P|s_2,s_1)
\;= \; B\bigl(\fr{Q}{2}+iP|s_2,s_1\bigr).
\]
\section{Partition function on the strip}
The naive definition $\Tr_{\CH^B}q^{H^B_{s_2,s_1}}$ is divergent.
It is better to consider {\bf relative} partition functions such as
\[ 
\CZ^{\rm B}_{s_2,s_1}(q)
=\Tr_{\CH^{\rm B}}\Bigl(q^{H^{\rm B}_{s_2,s_1}}-
q^{H^{\rm B}_{s_0,s_0}}\Bigr)
,\]
where $s_0$ parametrizes some fixed reference boundary condition.
This object provides information on the dependence of
the spectrum w.r.t. the boundary conditions. 
In view of \rf{spec} one expects $\CZ^{\rm B}$ to have the 
general form 
\begin{equation}\label{partfct2}
\CZ^{\rm B}_{s_2,s_1}(q)
=\int\limits_{\BS^{\rm B}}d\al \;\,\chi_{\al}(q)\;N(\al|s_2,s_1),
\end{equation}
where the Virasoro character $\chi_{\al}(q)$ is given by
\[ \begin{aligned}
\chi_{\al}(q)=& \eta^{-1}(q)q^{-(\al-\frac{Q}{2})^2}\\
=& q^{\frac{1-c}{24}+\De_{\al}}\prod_{k=1}^{\infty}(1-q^k).
\end{aligned}\]

Let us now present two guesses concerning the ingredients
of \rf{partfct2} that will later be confirmed independently:

First note that in the case
$\rho_2>0$ ,$\rho_1>0$ one would expect
the boundary contributions to the
Hamiltonian to be positive. In this case one does 
not expect any bound states:
\begin{equation}\label{specCardy} \BS^{\rm B}=\BS\equiv
\frac{Q}{2}+i\BR^+.
\end{equation}
Second, Al.B. Zamolodchikov has proposed 
\cite{Z} that the relation between 
spectral density $N$ and reflection amplitude $R$ that is 
well-known for quantum mechanical problems (see e.g. \cite{BY}, 
and \cite{MO} for a simple heuristic argument) will still work in the
present situation:
\begin{equation}\label{specshift} N(\be|s_2,s_1)\;=\;
\frac{1}{2\pi i N_0}\frac{\pa}{\pa P} \log \frac{R(P|s_2,s_1)}{
R(P|s_{\rm 0},s_{\rm 0})},
\end{equation}
where $\be$ and $P$ are related by $\be=\frac{Q}{2}+iP$.

\section{Cardy condition}
Consider Liouville theory on an annulus. 
There are two ways to describe the same 
amplitude: 
\begin{itemize} \item[$\bullet$]
as {\bf partition function} on the strip (periodic imaginary time)
\[ 
\CZ^{\rm B}_{s_2,s_1}(q)
=\int\limits_{\BS^B}d\be \;\,\chi_{\be}(q)\;N(\be|s_2,s_1).
\]
\item[$\bullet$]
as {\bf transition amplitude} in the theory with periodic boundary conditions.:
\begin{equation*}\label{cl_ch}
\begin{aligned}
{}_B^{}\bra s_2| & \tilde{q}^{H-\frac{c}{24}}|s_1\ket_B^{}-
{}_B^{}\bra s_0|\tilde{q}^{H-\frac{c}{24}}|s_0\ket_B^{}=\\[1ex]
= & \int\limits_{\BS}\!\frac{d\al}{2\pi}\;
\Bigl(  \bigl(A(\al|s_2)\bigr)^* A(\al|s_1)\\[-2ex]
 & \qquad
 -\,\bigl(A(\al|s_0)\bigr)^* A(\al|s_0)\Bigr)\chi_{\al}(\tilde{q}).
\end{aligned}\end{equation*} 
\end{itemize}
The Cardy condition will then be the relation
\begin{equation}\label{Cardcond}\boxed{
\begin{aligned} {}  \CZ^{\rm B}_{s_2,s_1}(q)
=& {}_B^{}\bra s_2|\tilde{q}^{H-\frac{c}{24}}|s_1\ket_B^{}\\
& -{}_B^{}\bra s_0|\tilde{q}^{H-\frac{c}{24}}|s_0\ket_B^{},
\end{aligned}}
\end{equation} 
where $\tilde{q}=\exp(-2\pi i /\tau)$
if $q=\exp(2\pi i\tau)$.

\subsection{Case a)}
In this case the validity of 
some analogue of the Cardy condition \rf{Cardcond}
was first verified by Al.B. Zamolodchikov \cite{Z}\footnote{From
the information available to the author it seems that
a different regularization was used in \cite{Z}.}. 
One may start with
the right hand side of \rf{Cardcond}. The
characters $\chi_{\al}(\tilde{q})$ can be expressed as sum
over charaters $\chi_{\al}(q)$ by means of  
\begin{equation}\label{modtr}
\chi^{}_{\al}(\tilde{q})=\int\limits_{\BS}d\be
\;\,S(\al,\be)\; 
\chi^{}_{\be}(q), 
\end{equation}
where  
\[ S(\al,\be)\;=
\;2\sqrt{2}\cos\bigl(4\pi (\al-\fr{Q}{2})(\al'-\fr{Q}{2})\bigr)
\]
In the presently 
considered case there is no 
problem to exchange the orders of integrations over 
$\al$ and $\be$:
\[ \begin{aligned}
{}_B^{} \bra s_2| & \tilde{q}^{H-\frac{c}{24}}|s_1\ket_B^{}\;=\\[1ex]
& \int_{\BS} d\be \;
\biggl\{\int_{\BS} d\al M(\al,\be|s_2,s_1)
\biggr\} \chi^{}_{\be}(q)
\end{aligned}
\]
where
\[ \begin{aligned}
M(\al,\be|s_2,s_1)=\;
\Bigl(  & \bigl(A(\al|s_2)\bigr)^* 
A(\al|s_1)\\[-1ex]
 - & \bigl(A(\al|s_0)\bigr)^* A(\al|s_0)\Bigr)\,S(\al,\be)
\end{aligned} \]
The integral in curly brackets can now easlily be identified 
with the integral representation
for $N(\be|s_2,s_1)$ that follows from the expression \cite{FZZ},
equation (3.18) for $B(\be|s_2,s_1)$ if $A_0$ and $N_0$ are related by
\[ \frac{1}{\pi N_0}\;=\;\sqrt{2}|A_0|^2. \]

\subsection{Case b)} 
Exchange of orders of integration produces 
a divergent result. Instead one may 
write the modular transformation of characters in the
form 
\[
\chi^{}_{\al}(\tilde{q})=\sqrt{2}
\int\limits_{r+i\BR}\!\!\!\!d\be
\;\,e^{4\pi i (\al-\frac{Q}{2})(\al'-\frac{Q}{2})}\; 
\chi^{}_{\be}(q), 
\] 
where the contour of integration was shifted into the 
half plane by an amount that will be chosen such that
\[
r>{\rm max}\bigl(|\si_+|,|\si_-|,\fr{Q}{2}\bigr)-\fr{Q}{2},
\]
where $\si_{\pm}\equiv i(s_2\pm s_1)\in\BR$.
Now it is possible to exchange orders 
of integrations over $\al$ and $\be$:
\[
\begin{aligned}  
{}_B^{}\bra  & s_2|\tilde{q}^{H-\frac{c}{24}}|s_1\ket_B^{}
 -{}_B^{}\bra s_0|\tilde{q}^{H-\frac{c}{24}}|s_0\ket_B^{}=\\
& = 
\;\int\limits_{r+i\BR} d\be \; N(\be|s_2,s_1)\;\chi^{}_{\be}(q) 
\end{aligned}\]
The contour of integration may be deformed into $\frac{Q}{2}+i\BR$ 
plus a finite sum 
of circles around poles of $N$. 
By using the fact that $S_b(x)$ has poles for $x=-nb-mb^-1$, 
$n,m\in\BZ^{\geq 0}$ \cite{Sh,FKV} 
one may now read off the spectrum on the strip:
\[
\boxed{\quad
\CH^{\rm B}\; = \;\int\limits_{\BS}d\al \;\, \CV_{\al}\;\;
\oplus\;\;\bigoplus_{\al\in\BD_{s_2s_1}}\;\,\CV_{\al},\quad}
\]
where 
\[
\begin{aligned}
\BD_{s_2s_1}=  \;\bigl\{ \be\in\BC\;|& \; \be=Q-|\si_{s}|+
nb+m\fr{1}{b}<\fr{Q}{2},\\
 & \text{ where }n,m\in\BZ^{\geq 0}, s=+,-\bigr\}.
\end{aligned} \]
By means of this calculation we have determined the
spectrum on the strip from the spectrum of Liouville theory on the cylinder!
Let us note that:\\[1ex]
{\bf 1.)} The spectrum is unitary only if $|\si_{\pm}|<Q$. Otherwise one 
finds nonunitary representations in the spectrum.\\[1ex]
{\bf 2.)} As long as both $|\si_{\pm}|<Q/2$ one still has
$\BD_{s_2s_1}=\emptyset$.\\[1ex]
{\bf 3.)} It is interesting, but puzzling to observe that there is a third case
where one obtains a spectrum that is compatible with hermiticity of the
Hamiltonian, namely $(s_1)^*=-s_2$. As noted earlier, one also finds
unitarity of the reflection amplitude in this case.

\section{Outlook: Connection to 
noncompact quantum group}

In order to carry out the program to determine the structure functions
as solutions of the consistency conditions they satisfy, it would be 
certainly quite useful if one could establish connections between
the structure functions and the characterisic data of (a generalization of)
a modular functor, such as e.g. the 
6j- or Racah-Wigner symbols of a quantum group and the 
associated modular transformation coefficients $S$. 
In the case of rational conformal field theories
connections of this kind have been exploited in \cite{BPPZ}\cite{FFFS} to 
derive expressions for the structure functions in terms of these data 
and to prove validity of the consistency equations on the basis of 
identities that the 
defining data of the modular functor have to satisfy. 

From this point of view the following two partial results look quite
encouraging: We have found (details will be given elsewhere) that
\begin{equation}\label{b-threept}
\boxed{C\BOC{\be_3}{s_3}{\be_2}{s_2}{\be_1}{s_1}\;=\;
M(\be_3,\be_2,\be_1)\SJSP{\si_1}{\be_1}{\be_2}{\si_3}{\si_2}{\be_3},
}
\end{equation}
where $\si_i\equiv is_i$, $\{\dots\}'_b$ are b-Racah-Wigner symbols associated to a
continuous series of representations of $\CU_q(\fsl(2,\BR))$.
They differ from those constructed in \cite{PT2} by a change of normalization
of the Clebsch-Gordan coefficients:
\[ \begin{aligned}
{} &\SJSP{\al_1}{\al_2}{\al_3}{\al_4}{\al_s}{\al_t}=\\
& \frac{S_b(\al_t+\al_4-\al_1)S_b(\al_3+\al_{t}-\al_2)}
{S_b(\al_4+\al_3-\al_s)S_b(\al_s+\al_2-\al_1)}
\SJS{\al_1}{\al_2}{\al_3}{\al_4}{\al_s}{\al_t}.
\end{aligned}\]

The expression \rf{b-threept} 
for the boundary three point function satisfies the 
associativity condition \rf{b-ass} due to the relation between
fusion coefficients $F$ and b-Racah-Wigner symbols \cite{PT1}, as
well as the pentagon equation satisfied by the latter \cite{PT2}. 
Condition \rf{b-ass} clearly leaves the freedom to
change the normalization of the boundary fields. This freedom
can be fixed by considering special cases of \rf{b-ass} where
one of $\be_4,\dots,\be_1$ is chosen to correspond to the 
degenerate representation $\CV_{-b}$. One then obtains
{\it linear} finite difference equations for the 
boundary three point function with coefficients given in terms
of know fusion coefficients and the special three point function
$c_-$ calculated in \cite{FZZ}\footnote{The evaluation of the 
integral for $c_-$ that was not presented in \cite{FZZ} may be circumvented
by using old results of Gervais and Neveu \cite{GN}}.

It seems moreover suggestive to observe that the formula for the 
one point function can for $\al\in\BS$ be rewritten in a form that
resembles the Cardy formula \cite{C} for the one-point function:
\begin{equation}\label{onept}
\boxed{\quad A(\al|s)\;=\;e^{i\de(\al)}
\frac{S(\al;s)}{\sqrt{\mu(\al)}},\qquad}
\end{equation}
where $e^{2i\de(\al)}\equiv R(\al)$ is the bulk reflection
amplitude \cite{ZZ}, and $\mu(\al)$ turns out to be the
Plancherel measure of the quantum group dual to the 
category of representations of $\CU_q(\fsl(2,\BR))$ that
was considered in \cite{PT1}\cite{PT2}. One clearly sees
in what respects the bootstrap in our noncompact case is 
richer than in RCFT (reflection amplitude, loss of direct relation
between ``quantum dimension'' and modular transformation coefficients).
On the other hand one may well consider \rf{onept} to be 
the most natural generalization of the Cardy formula for the 
one point function that one may reasonably hope for.  

\acknowledgments 
This work develops further some results and
suggestions that have been presented by Al. B. Zamolodchikov
in \cite{Z}. The author would like to thank him for very
inspiring discussions.

\end{document}